\documentclass[aps,prb,preprint,groupedaddress]{revtex4}

\usepackage{graphicx}
\usepackage{dcolumn}
\usepackage{bm}
\usepackage{amssymb}
\usepackage{ulem}
\usepackage{xcolor}
\usepackage{marginnote}
\usepackage{color}
\usepackage{soul}
\usepackage{graphicx}
\usepackage{amsmath}
\usepackage{amssymb}
\usepackage{bm}
\usepackage{dcolumn}
\usepackage{longtable}
\usepackage{multirow}

\begin{document}


\title{Resonant enhancement of the near-field radiative heat transfer in nanoparticles}


\author{S.~G.~Castillo-L\'opez}
\affiliation{Instituto de F\'{i}sica, Universidad Nacional Aut\'onoma de M\'exico, Apartado Postal 20‐364, M\'exico 01000, M\'exico.}
\author{A. Márquez}
\affiliation{Instituto de F\'{i}sica, Universidad Nacional Aut\'onoma de M\'exico, Apartado Postal 20‐364, M\'exico 01000, M\'exico.}
\author{R. Esquivel-Sirvent}
\affiliation{Instituto de F\'{i}sica, Universidad Nacional Aut\'onoma de M\'exico, Apartado Postal 20‐364, M\'exico 01000, M\'exico.}



\date{\today}

\begin{abstract}
We numerically study the tuning of the radiative heat transfer between a spherical InSb nanoparticle in the vicinity of a flat SiC surface assisted by a static magnetic field.  By changing the value of the applied magnetic field, the dielectric function of the nanosphere becomes anisotropic due to the excitation of magneto-plasmons.  In the dipolar approximation, the plasmon resonance of the particle splits into two additional satellite resonances that shift to higher and lower frequencies as the field increases.
When one of the particle resonances overlaps with the phonon-polariton frequency of the SiC surface, an enhancement of the heat transfer of two orders of magnitude is obtained. 
To understand the tuning of the radiative heat transfer, we present a detailed analysis of the nature of the modes that can be excited (surface, bulk, and hyperbolic). 
\end{abstract}
%

\maketitle

\section{Introduction}

The near-field radiative heat transfer (NFRHT) between two bodies at different temperatures is characterized by an excess heat flux larger than that predicted by the Stefan-Boltzmann law for black bodies. Due to the many potential applications of the NFRHT at the micro and nanoscale, there has been intense research activity in the field \cite{PhysRevLett.112.044301,desutter2019near,marconot2021toward}.
Precise experimental measurements of the heat flux  at submicron separations are now commonplace \cite{Kittel05,Francoeur15,Dewilde,Gelais,kittel}. The theoretical explanation for the NFRHT is based on the fluctuation-dissipation theorem and Rytov's theory of thermally excited electromagnetic fields in materials \cite{Vinogradov}. Thus, the dielectric function and magnetic susceptibility play an important role in defining the heat flux \cite{Vinogradov,vanhove,HARGREAVES}.  

\noindent
The dependence of the heat flux on the dielectric function of the medium opens the possibility of tuning or controlling its radiative properties. Modification of the dielectric response can be achieved in several ways. In composite materials, the dielectric function can be modified by a suitable combination of host and inclusions \cite{santiago2017dispersive,santiago2017near}. The simplest configuration of a composite is a layered media that can give rise to the different surface and hyperbolic modes \cite{biehs2007thermal,ben2009near,Thinfilmraul,JaimeACS,
lim2018tailoring}, which allow the enhancement of the radiative heat transfer at subwavelength scales \cite{PhysRevMaterials.3.015201,Esquivel17}.
The dielectric function can also change during a phase transition, so materials such as VO$_2$ are useful for modulating the total heat flux in the near field \cite{ghanekar2018near}.
A similar result was predicted using YBCO superconductors that show a drastic modification of their dielectric response with temperature \cite{castillo2020near,castillo2022enhancing}.  

Another system of interest is nanoparticles: either the case of the heat transfer between two or more particles \cite{PhysRevB.99.045418,Wang16,chapuis2008radiative} or between a substrate and a nanoparticle \cite{bai2015enhanced,huth2010shape}. In these systems, the NFRHT is determined by the absorption cross-section of the nanoparticle, which depends not only on dielectric function but also on the shape of the particles \cite{shape1,Cecilia}.  Thus, the geometry also becomes an important parameter to tune the NFRHT. The case of heat transfer between a plane and spheroidal nanoparticles shows that the heat flux can be tuned by changing the aspect ratio \cite{huth2010shape}.

Another possibility is to have a spherical nanoparticle with an anisotropic dielectric function. This can be achieved through the excitation of magneto-plasmons (MPs), which arise from the interaction of a localized plasmon with an external magnetic field  \cite{KUSHWAHA20011}.
Historically, the most common doped semiconductor in which magneto-plasmons have been observed is InSb, because small magnetic fields are needed \cite{keyes1956infrared,palik1961infrared,Palik_1970}. Furthermore, since MPs are  in the THz region they are  suitable for  optical applications \cite{chochol2016magneto,dragoman2008plasmonics}. The  excitation of magneto-plasmons depends also on the shape of the nanoparticles. In the work of Pedersen \cite{PhysRevB.102.075410} a detailed study of magneto-plasmon resonances in nanoparticles with different shapes was presented. 

The NFRHT  between two parallel plates with a constant applied magnetic field can be used to control the heat transfer. For doped semiconductors, the heat flux can be reduced by about 700\%. 
In the case of nanoparticles it is also possible to excite magneto-plasmons \cite{zhang2015surface,InSbparticle}. Magneto-plasmonic effects were considered as means of controlling the heat flux in an array of nanospheres \cite{abraham2018anisotropic} and inducing a giant  magnetoresistance as a function of the applied magnetic field \cite{PhysRevLett.118.173902}. These applications involved InSb particles, since it is the doped semiconductor that exhibits a magneto-plasmonic response at relatively small magnetic fields \cite{InSbparticle}. The anisotropy of the dielectric function of the nanoparticle will have an effect on the polarizability that plays an important role in the calculation of the NFRHT. The sensitivity to the external magnetic field of some nanostructures has been demonstrated in the fine splitting of their plasmonic resonances \cite{Marquez:20,PhysRevLett.126.136801}.

In this work, we calculate  the NFRHT between a SiC plane and an InSb spherical nanoparticle \cite{kordesch} in the presence of an external magnetic field that induces an anisotropy in the dielectric function. We find an increase in the total heat flux of up to two orders of magnitude when the maximum peak of the nanoparticle polarizability is tuned to the SiC surface-phonon resonance.

\section{Dielectric functions and polarizability} 

\begin{figure}[h!]
	\includegraphics[width=0.5\textwidth]{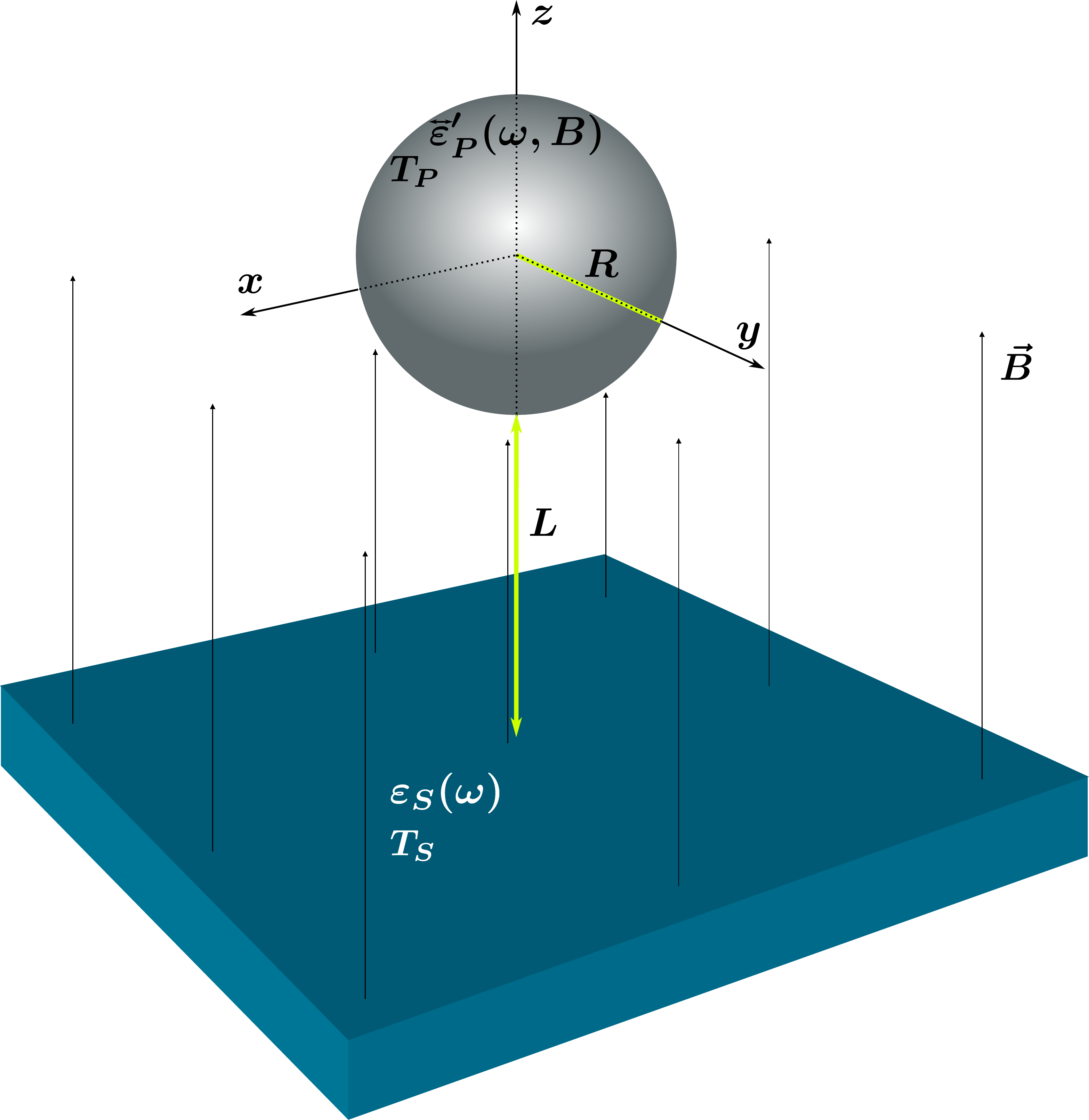}
	\caption{Sketch of the system: Nanoparticle with radius $R$, permitivity $\tensor{\varepsilon}'_P(\omega,B)$, and temperature $T_P$, located at a distance $L$ from a planar substrate of nonmagnetic material with permitivity $\varepsilon_{S}(\omega)$, and temperature $T_S$. The applied constant magnetic field, $\vec{B}$, is directed along the $z$-axis.}
	\label{fig-01}
\end{figure}

We consider the system depicted in Fig.~\ref{fig-01} that consist of a SiC surface with a dielectric function $\varepsilon_S(\omega)$, and an InSb sphere of radius $R$ with a dielectric function $\varepsilon_P(\omega)$. The separation of the particle and the plane is $L$, which we assume to be vacuum.  
When an external magnetic field is applied, the dielectric response of the nanoparticle becomes anisotropic~\cite{Palik_1970,palik1961infrared,PhysRevB.12.3186,Weick}. The diagonalized dielectric tensor is described by  $\tensor{\varepsilon}_P'(\omega,B)=diag\left(\varepsilon_{xx}+i\varepsilon_{xy},~\varepsilon_{xx}-i\varepsilon_{xy},~\varepsilon_{zz}\right)=diag\left(\varepsilon_{P,x}',\varepsilon_{P,y}',\varepsilon_{P,z}' \right)$, where the components have an electronic and a phononic contribution given as:  
\begin{subequations}\label{eq:eps_InSb}
	\begin{align}
		\varepsilon_{xx}(\omega,B)=\varepsilon_{\infty,P}&\left\{1-\frac{(\omega+i\gamma)\omega_p^2}{\omega\left[(\omega+i\gamma)^2-\omega_c^2\right]} +\right.\\[4pt]\nonumber
		&\left.\frac{\omega_{L,P}^2-\omega_{T,P}^2}{\omega_{T,P}^2-\omega^2-i\Gamma_P\omega}\right\},
	\end{align}
	\begin{equation}
		\varepsilon_{xy}(\omega,B)=i\varepsilon_{\infty,P}\left\{\frac{\omega_c\omega_p^2}{\omega\left[(\omega+i\gamma)^2-\omega_c^2\right]}\right\},
	\end{equation}
	\begin{equation}
		\varepsilon_{zz}(\omega)=\varepsilon_{\infty,P}\left[1-\frac{\omega_p^2}{\omega(\omega+i\gamma)} + \frac{\omega_{L,P}^2-\omega_{T,P}^2}{\omega_{T,P}^2-\omega^2-i\Gamma_P\omega}\right].
	\end{equation}
\end{subequations}
The cyclotron frequency is  $\omega_c(B)=eB/m^*$ written in terms of the effective mass $m^*$. 
For  InSb, the parameters are $\varepsilon_{\infty,P}=15.7$, $\omega_p=0.314~\omega_0$, $\gamma=0.034~\omega_0$, and $m^*=0.22m_e$.
Phonon parameters are $\omega_{T,P}=0.339~\omega_0$, $\omega_{L,P}=0.362~\omega_0$, and $\Gamma_P=5.65\times 10^{-3}~\omega_0$.
Throughout the paper, all the frequencies are normalized to $\omega_0=10^{14}$ rad/s.
In the case of $B=0$ the dielectric function becomes isotropic since $\omega_c=0$. 

For small nanoparticles where higher order multipoles can be ignored, the power absorbed by a nanoparticle depends on the magnetic and electric  dipole  contribution. The electric and magnetic polarizabilities are defined, respectively, as  

\begin{equation}\label{eq:alpha_E}
		\alpha_{i}^E=4\pi R^3\frac{\varepsilon_{P,i}'(\omega)-1}{\varepsilon_{P,i}'(\omega)+2},
\end{equation}
\begin{equation}\label{eq:alpha_H}
		\alpha_{i}^H=\frac{2\pi}{15}R^3(k_0 R)^2\left[\varepsilon_{P,i}'(\omega)-1\right],\\
\end{equation}
where the index $i=x,y,z$. 

The dielectric function  $\varepsilon_{S}(\omega)$ of SiC is described by the expression 
\begin{equation}
\varepsilon_{S}(\omega)=\varepsilon_{\infty,S}\left(1+\frac{\omega_{L,S}^2-\omega_{T,S}^2}{\omega_{T,S}^2-\omega^2-i\Gamma_S\omega}\right),
\end{equation}
where $\varepsilon_{\infty,S}= 6.7$, $\omega_{L,S}=1.825~\omega_0$, $\omega_{T,S}=1.493~\omega_0$, and $\Gamma_S=8.97\times10^{-3}~\omega_0$.  

\section{NFRHT equations} 

When the characteristic thermal wavelength, $\lambda_T=\hbar c/ k_B T$, of the radiation emitted by the substrate is larger than the radius $R$ of the particle, it can be considered as a dipole whose electromagnetic response is described by the electric (\ref{eq:alpha_E}) and magnetic (\ref{eq:alpha_H}) polarizabilities.   

Then, within the framework of fluctuating electrodynamics theory, the total heat flux, $Q_T$, exchanged by an anisotropic spherical particle at temperature, $T_P$, separated a distance, $L$, from a semi-infinite substrate at temperature, $T_S$, can be calculated as follows \cite{huth2010shape}, 
\begin{equation}\label{eq:QT}
	Q_T(T_S,T_P,L)=\int_{0}^{\infty}d\omega S_{\omega}(\omega,T_S,T_P,L).
\end{equation}
The spectral heat flux, $S_{\omega}$, is given by the expression:
\begin{align}\label{eq:Sw}
	S_{\omega}(\omega,T_S,T_P,L)&=8\left[\Theta(\omega,T_S)-\Theta(\omega,T_P)\right]k_0^2\times\\[4pt]\nonumber
	&\int \frac{d\vec{\boldsymbol{\kappa}}}{(2\pi)^3}\left[\tau^{prop}(\omega,\kappa,L)+\tau^{evan}(\omega,\kappa,L)\right],
\end{align}
where $\Theta(\omega,T)=\hbar\omega/(\exp\left(\hbar\omega/k_B T\right)-1)$ is the Planckian distribution, $k_0=\omega/c$, and $\vec{\boldsymbol{\kappa}}=(k_x,k_y)=(\kappa\cos\phi,\kappa\sin\phi)$ is the longitudinal wavevector, parallel to the substrate interface. 
The contributions of propagating and evanescent waves to the heat flux are determined by the coefficients $\tau^{prop}(\omega,\kappa,L)$ and $\tau^{evan}(\omega,\kappa,L)$, respectively, which are deduced from Ref. ~\cite{huth2010shape}. These quantities give the mean energy distribution resolved in frequency ($\omega$)-wavevector ($\kappa$) space:
\begin{subequations}\label{eq:Integrando-wkappa}
\begin{align}\label{eq:Integrando-wkappa-prop}
	\tau^{prop}&(\omega,\kappa,L)=\frac{1}{4k_{z0}}\Bigg\{\\\nonumber
	&\mbox{Im}(\alpha_{x}^E)\left[\sin^2\phi (1-|r_{s}|^2)+\frac{\cos^2\phi k_{z0}^2}{k_0^2}(1-|r_{p}|^2)\right]+\\\nonumber
	&\mbox{Im}(\alpha_{y}^E)\left[\cos^2\phi (1-|r_{s}|^2)+\frac{\sin^2\phi k_{z0}^2}{k_0^2}(1-|r_{p}|^2)\right]+\\\nonumber
	&\mbox{Im}(\alpha_{z}^E)\frac{\kappa^2}{k_0^2}(1-|r_{p}|^2)+\\\nonumber
	&\mbox{Im}(\alpha_{x}^H)\left[\sin^2\phi (1-|r_{p}|^2)+\frac{\cos^2\phi k_{z0}^2}{k_0^2}(1-|r_{s}|^2)\right]+\\\nonumber
	&\mbox{Im}(\alpha_{y}^H)\left[\cos^2\phi (1-|r_{p}|^2)+\frac{\sin^2\phi k_{z0}^2}{k_0^2}(1-|r_{s}|^2)\right]+\\\nonumber
	&\mbox{Im}(\alpha_{z}^H)\frac{\kappa^2}{k_0^2}(1-|r_{s}|^2)\\\nonumber
	&\Bigg\},\qquad \mbox{for} \quad k_0>\kappa,
\end{align}
and
\begin{align}\label{eq:Integrando-wkappa-evan}
	\tau^{evan}&(\omega,\kappa,L)=\frac{\exp(-2\gamma L)}{2\gamma}\Bigg\{\\\nonumber
	&\mbox{Im}(\alpha_{x}^E)\left[\sin^2\phi \mbox{Im}(r_{s})+\frac{\cos^2\phi \gamma^2}{k_0^2}\mbox{Im}(r_{p})\right]+\\\nonumber
	&\mbox{Im}(\alpha_{y}^E)\left[\cos^2\phi \mbox{Im}(r_{s})+\frac{\sin^2\phi \gamma^2}{k_0^2}\mbox{Im}(r_{p})\right]+\\\nonumber
	&\mbox{Im}(\alpha_{z}^E)\frac{\kappa^2}{k_0^2}\mbox{Im}(r_{p})+\\\nonumber
	&\mbox{Im}(\alpha_{x}^H)\left[\sin^2\phi \mbox{Im}(r_{p})+\frac{\cos^2\phi \gamma^2}{k_0^2}\mbox{Im}(r_{s})\right]+\\\nonumber
	&\mbox{Im}(\alpha_{y}^H)\left[\cos^2\phi \mbox{Im}(r_{p})+\frac{\sin^2\phi \gamma^2}{k_0^2}\mbox{Im}(r_{s})\right]+\\\nonumber
	&\mbox{Im}(\alpha_{z}^H)\frac{\kappa^2}{k_0^2}\mbox{Im}(r_{s})\\\nonumber
	&\Bigg\},\qquad \mbox{for} \quad k_0<\kappa.
\end{align}	
\end{subequations}

The above expressions are in terms of the different components of the particle polarizability tensors $\alpha^E_{i}$ and $\alpha^H_{i}$, and also in terms of the reflection coefficients $r_s$ and $r_p$ associated with the incidence of $p$ and $s$-polarized light on the substrate, respectively. These coefficients are given by the well-known Fresnel formulas:
\begin{equation}
	r_s=\frac{k_{z0}-k_z}{k_{z0}+k_z},\qquad \mbox{and}\qquad r_p=\frac{\varepsilon_{S}(\omega)k_{z0}-k_z}{\varepsilon_{S}(\omega)k_{z0}+k_z}.
\end{equation}
Here $k_{z0}=\sqrt{k_0^2-\kappa^2}=i\gamma$ and $	k_{z}=\sqrt{\varepsilon_{S}(\omega)k_0^2-\kappa^2}$ are the transversal wavevectors normal to the interface (vacuum|substrate) that correspond to the vacuum and the material space of permittivity $\varepsilon_{S}$, respectively.

From equations (\ref{eq:QT})-(\ref{eq:Integrando-wkappa}), we  obtain the expressions for the radiative heat transfer between an interface and a spherical nanoparticle made of an isotropic material by considering $\alpha_{x}^{E,H}=\alpha_{y}^{E,H}=\alpha_{z}^{E,H}=\alpha^{E,H}$.  
For the case of the evanescent field, the formulas are presented in Ref.~\cite{PhysRevB.77.125402}, Ecs. (3-10,16). 
On the other hand, the part associated with propagating waves conduces to the textbook result $Q_T=1/\pi^2\int_{0}^{\infty}d\omega\Theta(\omega,T_P)k_0^3\mathrm{Im}(\alpha^E)$ for the energy emitted by a nanoparticle of a nonmagnetic material in absence of external surfaces ($r_s=r_p=0$), see Ref.~\cite{bohren2008absorption}, pp. 124,140.

\section{Results}
We conducted numerical experiments of the NFRHT considering an InSb nanoparticle with radius $R=20$ nm and temperature $T_{P}=100$ K placed at $L=50$ nm above a planar SiC substrate with temperature $T_{S}=300$ K. The system is depicted in Fig.~\ref{fig-01}.

\begin{figure}[h!]
	\includegraphics[width=0.5\textwidth]{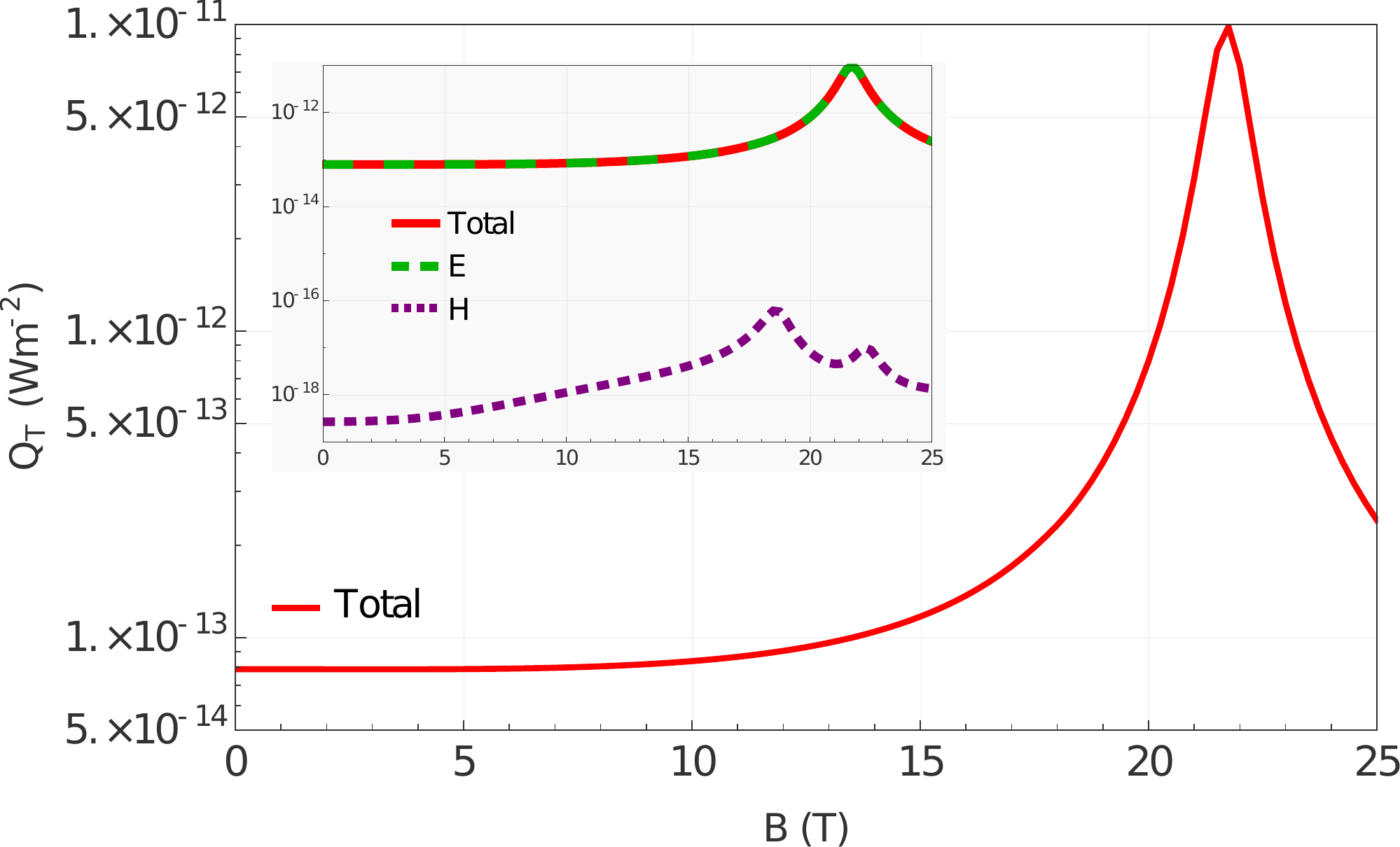}
	\caption{Total heat flux, $Q_T$, as a function of the applied constant magnetic field, $B$, for the setup depicted in Fig.\ref{fig-01}. The system consists of an InSb particle with radius $R=20$ nm and temperature $T_{P}=100$ K, separated by $L=50$ nm from a SiC substrate with temperature $T_{S}=300$ K. Electric and magnetic contributions to $Q_T$ are shown in the figure inset.}
	\label{fig-02}
\end{figure}

Based on Eqs.~(\ref{eq:QT})-(\ref{eq:Integrando-wkappa}), we plot in Fig.~\ref{fig-02} the total heat flux, $Q_T$, as a function of the magnitude of the applied field, $B$.
For  small values of the magnetic field $B$, the total heat flux shows small  changes. However, when the magnetic field exceeds approximately $10$ T, the heat flux increases at a faster pace, until it reaches its maximum value around $B=21.75$ T. 
This optimal value of $B$,  enhances the NFRHT of the system by two orders of magnitude in comparison with the case without magnetic field, $B=0$.
When the field exceeds the optimal value, the heat flux drops again, suggesting a resonant behavior.
In the inset of Fig.~\ref{fig-02}, the electric and magnetic contributions to the total heat flux are shown as the green-dashed and purple-dotted curves, respectively. It demonstrates that the energy transfer is totally governed by 
the electric polarizability of the InSb nanoparticle, $\alpha_{i}^E$. 

\begin{figure}[h!]
	\includegraphics[width=0.5\textwidth]{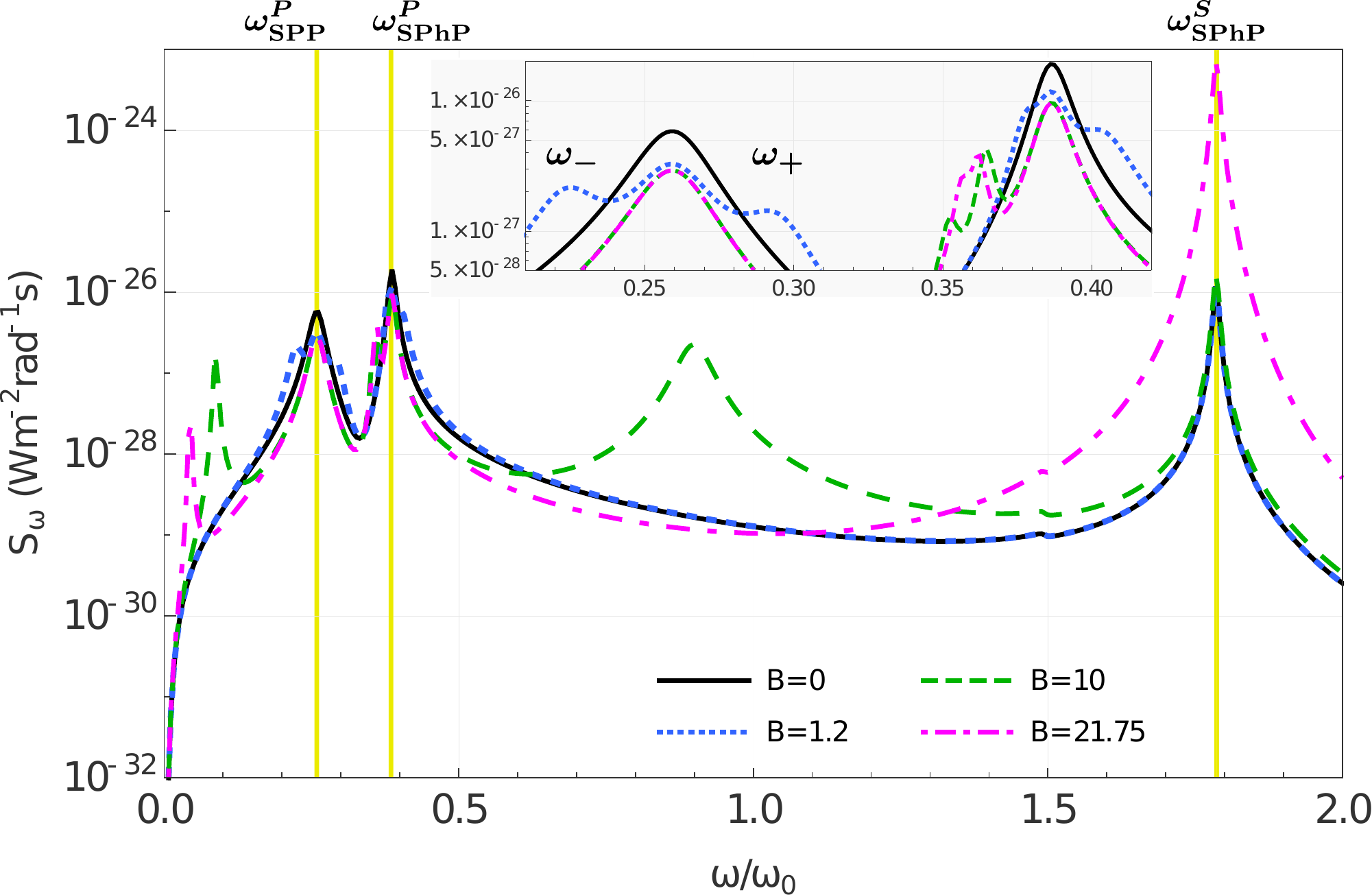}
	\caption{Spectral heat flux, $S_{\omega}$, between an InSb particle with radius $R=20$ nm and temperature $T_{P}=100$ K separated by $L=50$ nm vacuum gap from a SiC substrate with temperature $T_{S}=300$ K. The colors of the curves correspond to different values of the external magnetic field.}
	\label{fig-03}
\end{figure}

The origin of the two orders of magnitude increase in the total heat flux for a specific value of the applied magnetic field can be understood  by analyzing the spectral distribution of the transferred energy, $S_\omega$, given by Eq.~(\ref{eq:Sw}) and shown in Fig.~\ref{fig-03}. Here, the colors of the curves correspond to different values of the magnetic field indicated in the figure. 
In the absence of the magnetic field, the spectral heat flux is characterized by two peaks at low frequency: one at $\omega_{\mathrm{SPP}}^P=0.26~\omega_0$ associated with the surface plasmon-polariton (SPP) of the nanoparticle, and the other one at $\omega_{\mathrm{SPhP}}^P=0.39~\omega_0$ related to the excitation of surface phonon-polariton (SPhP) modes.
Completely decoupled from the previous two, the SPhP resonance of SiC is present at $\omega_{\mathrm{SPhP}}^S=1.78~\omega_0$.

By turning on the magnetic field, the original SPP resonance of the nanoparticle is split into two new satellite resonances maintaining the original one at $\omega_{\mathrm{SPP}}^P$. 
As a consequence, additional peaks $(\omega_-,\omega_+)$ are apparent in the heat flux spectra associated with the excitation of magneto-plasmons below $\omega<0.3~\omega_0$, see the inset of Fig.~\ref{fig-03}. 
For small magnitudes of the magnetic field, the frequency shift of the satellite resonances displays the linear dependence $\omega_{\pm}\approx\omega_{\mathrm{SPP}}^P\pm\omega_c/2$, as the curve $B=1.2$ T exemplifies. 
When the external field increases, nonlinear behavior is observed. This is clearly appreciated in the case $B=10$ T, where the two satellite peaks are no longer equidistant from the original plasmon resonance.
This behavior of the magneto-plasmons resembles the atomic Zeeman effect and is known as plasmonic Zeeman effect \cite{Marquez:20}. 
The high sensitivity to the magnetic field of the redshift resonance $\omega_+$ allows tuning its corresponding spectral heat flux peak to that associated with the SiC-SPhP by applying the optimal field $B\approx21.75$ T. 
In this case, the heat flux around the resonance frequency $\omega_{\mathrm{SPhP}}^S$ increases by almost three orders of magnitude.

The resonance associated with the nanoparticle surface phonon $\omega_{\mathrm{SPhP}}^P$ also exhibits a slight splitting due to the red/blue-shift of the satellite resonances of the SPP, but there is no explicit dependence of the phonon band on the magnetic field.
In fact, the magneto-optical response enters into the InSb dielectric response, Eq.~(\ref{eq:eps_InSb}), through the free charge carriers band via the cyclotron frequency, $\omega_c$. Thus, as B increases, $\omega_c$ carries the dispersion of the free charges to the lower and higher frequencies, $\omega_{\pm}$. This indirectly modifies the dispersion of the nanoparticle SPhP causing the appearance of two extra peaks in the frequency range $\omega\approx 0.35-0.42~\omega_0$.

\begin{figure*}[h!]
	\includegraphics[width=0.7\textwidth]{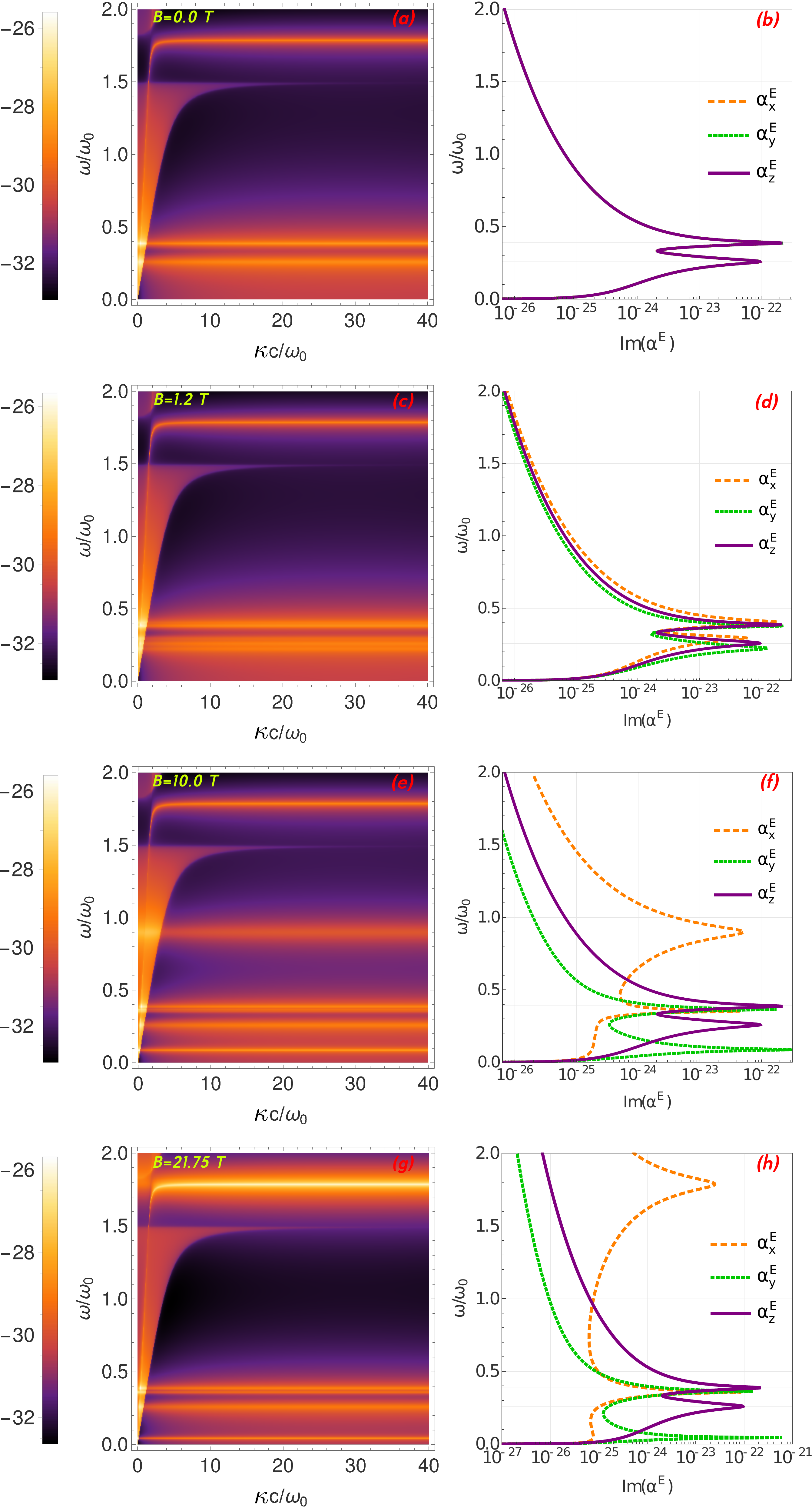}
	\caption{(a, c, e, g) Contribution of P-polarized waves to the mean energy distribution resolved in the normalized space of longitudinal wavevectors and frequencies, $\kappa-\omega$. (b, d, f, h) Imaginary part of the nanoparticle electric polarizability, Im$\alpha_{i}^E$, as a function of the frequency, $\omega/\omega_0$. In the figure, each row corresponds to a different value of the magnetic field, $B$.}
	\label{fig-04}
\end{figure*}

To better understand the nature of the different peaks in the heat flux spectrum, Figure~\ref{fig-04} shows the mean energy distribution (in logarithmic scale) throughout the $\kappa$-$\omega$ space for the case of P-polarized waves. Density-plots (a, c, e, g) were obtained using Eqs. (\ref{eq:Integrando-wkappa}), (\ref{eq:alpha_E}) and (\ref{eq:eps_InSb}) for different values of the external magnetic field. 
Figures (b, d, f, h), present the imaginary part of the electric polarizability, $\alpha_{i}^E$, as a function of the normalized frequency for the InSb nanoparticle immersed in the corresponding magnetic field.
In Fig.~\ref{fig-04}(a) the mean energy profile exhibits two maximum values at $\omega\approx\omega_{\mathrm{SPP}}^P=0.26~\omega_0$ and $\omega\approx\omega_{\mathrm{SPhP}}^P=0.39~\omega_0$ related to the InSb plasmon- and phonon-polariton resonances, respectively, while the SiC phonon-polariton contribution is observed at $\omega\approx\omega_{\mathrm{SPhP}}^S=1.78~\omega_0$. 
In this isotropic situation, the three components of the nanoparticle electric polarizability coincide having resonances at the frequencies $\omega_{\mathrm{SPP}}^P$ and $\omega_{\mathrm{SPhP}}^P$, see Fig. \ref{fig-04}(b).

Anisotropic dielectric response arises in the presence of a magnetic field: while the $z$-component of the electric polarizability is unchanged, the $x$- and $y$- components show magneto-plasmon resonances shifted to higher and lower frequencies than the original plasmon, respectively.
Where a magnetic field of $B=1.2$ T is applied, the new satellite plasmon resonances are almost equidistant from the original one, see Figs.~\ref{fig-04}(c,d).
Also, the InSb phonon-polariton splits into two extra resonances. As a result, six different nanoparticle modes are available for radiative energy transfer. 
From Figs.~\ref{fig-04}(e,f), we observed nonlinear shift of the magneto-plasmon resonances for the case $B=10.0$ T.
On the other hand, the resonance frequencies of satellite phonon-polariton modes are weakly influenced by the magnetic field in comparison with the behavior of plasmonic ones.
Figures.~\ref{fig-04}(g,h) illustrate the situation of maximum coupling between the nanoparticle satellite peak $\omega_{+}$ with the SiC resonance peak.

The magnetic field not only modifies the dispersion of the nanoparticle plasmonic and phononic surface modes, but it also changes their nature: In the absence of magnetic field, the resonance condition $\varepsilon_P(\omega)+2=0$ of the electric polarizability is associated with true surface modes (SMs). 
On the other hand, when the magnetic field is applied, the anisotropy of the dielectric function gives rise to the appearance of hyperbolic modes (HMs). They can propagate across the nanoparticle but decays in its surroundings. HMs are expected within the frequency regions where the real part of almost one component of the permittivity tensor $\tensor{\varepsilon}_P'$ has an opposite sign with respect to the other two~\cite{hong2020biaxial,song2018biaxial}, i.e., $\mbox{Re}(\varepsilon_{P,i}'),~\mbox{Re}(\varepsilon_{P,j}')<0,~\mbox{Re}(\varepsilon_{P,k}')>0$ or $\mbox{Re}(\varepsilon_{P,i}'),~\mbox{Re}(\varepsilon_{P,j}')>0,~\mbox{Re}(\varepsilon_{P,k}')<0$.

\begin{figure*}[h!]
	\includegraphics[width=0.9\textwidth]{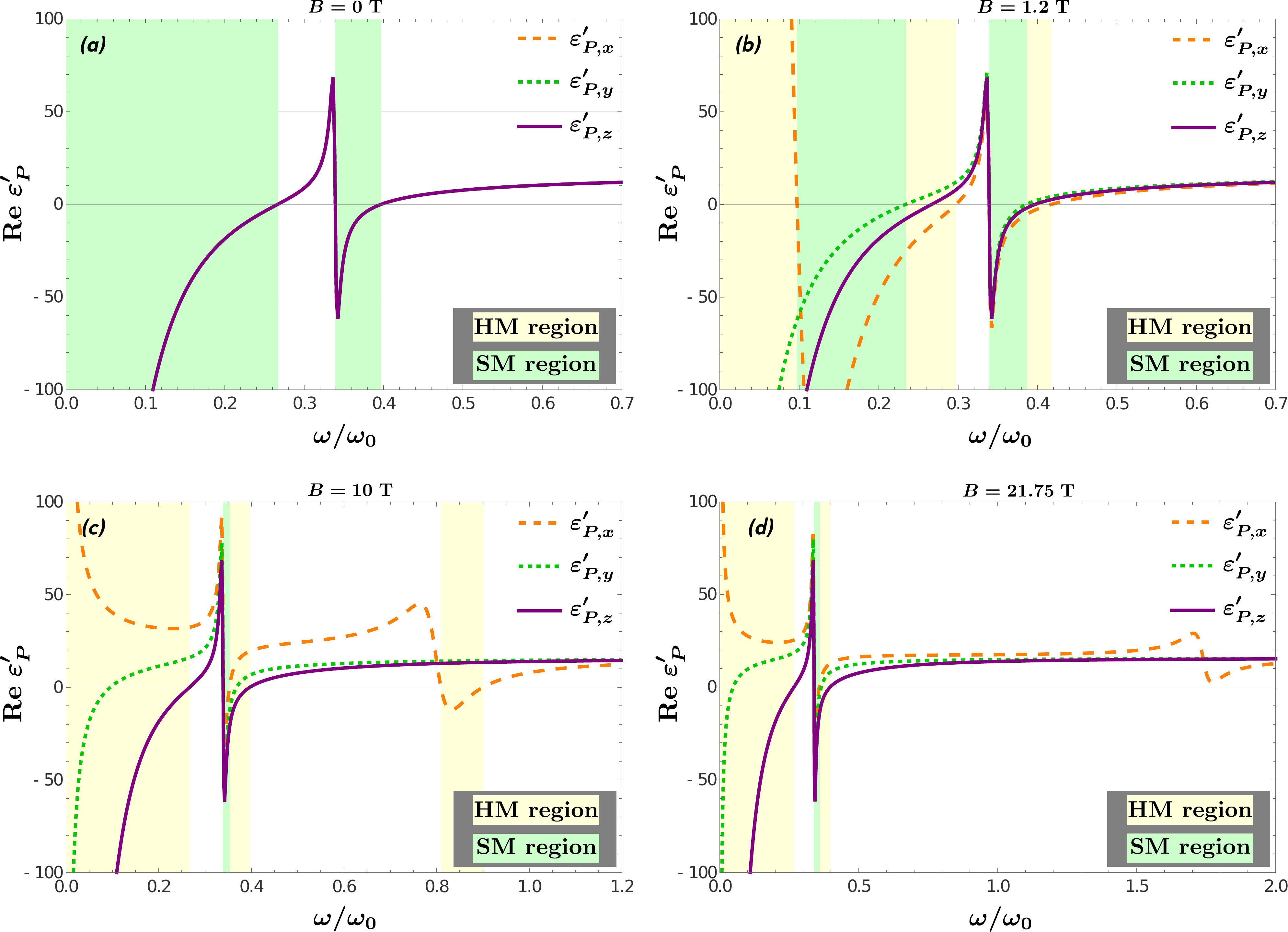}
	\caption{Components of the permittivity tensor  $\tensor{\varepsilon}_P'$ given by Eq. \ref{eq:eps_InSb}. The hypebolic mode (HM) regions correspond to the frequency range where  the real part of  one component of $\tensor{\varepsilon}_P'$ has an opposite sign with respect to the other two, i.e., $\mbox{Re}(\varepsilon_{P,i}'),~\mbox{Re}(\varepsilon_{P,j}')<0,~\mbox{Re}(\varepsilon_{P,k}')>0$ or $\mbox{Re}(\varepsilon_{P,i}'),~\mbox{Re}(\varepsilon_{P,j}')>0,~\mbox{Re}(\varepsilon_{P,k}')<0$. In the surface mode (SM) regions the three components of the permittivity tensor exhibit negative values, $\mbox{Re}(\varepsilon_{P,x}'),~\mbox{Re}(\varepsilon_{P,y}'),~\mbox{Re}(\varepsilon_{P,z}')<0$.}
	\label{fig-05}
\end{figure*}

Figure \ref{fig-05} shows the components of the InSb permitivity tensor. Here the frequency range of HMs corresponds to the yellow shaded regions, while the green shaded ones correspond to SMs, where the permitivity components satisfy $\mbox{Re}(\varepsilon_{P,x}'),~\mbox{Re}(\varepsilon_{P,y}'),~\mbox{Re}(\varepsilon_{P,z}')<0$.
When $B=1.2,T$ is applied, the original nanoparticle resonance at $\omega_{\mathrm{SPP}}^P$ and the upper magneto-plasmon resonance lies within the HM region $\omega\approx0.23-0.3~\omega_0$. Thus only the lower satellite resonance $\omega_{-}$ is actually a true SM, see Fig. \ref{fig-05}(b).
As the field magnitude increases, the HM regions widen while the SM regions begin to reduce.
For a field of $B=10$ T, the original plasmonic resonance and the two satellite ones are now associated with HMs. Only the SPhPs survive within the reststrahlen band as shown in Fig. \ref{fig-05}(c). 
By increasing the magnetic field, even more, it begins to destroy the HM regions too, see the case for $B=21.75$ T in Fig. \ref{fig-05}(d). 
Here, the magnetic field drastically modifies the dielectric response of the nanoparticle, disappearing the high-frequency magneto-plasmon $\omega_+$.
Thus the resonance condition $\tilde{\varepsilon}'_{P,x}(\omega)=-2$ is no longer satisfied, however, the imaginary part of the nanoparticle electric polarizability, $\mbox{Im} (\alpha_x^E)$, displays a maximum value at the frequency where this plasmon would be located, see Fig.~\ref{fig-04}(h). For $B=21.75$, this maximum peak coincides with the phonon resonance of SiC substrate, giving rise to an increase of the mean energy distribution we observed at $\omega\approx1.78~\omega_0$.

\begin{figure}[h!]
	\includegraphics[width=0.5\textwidth]{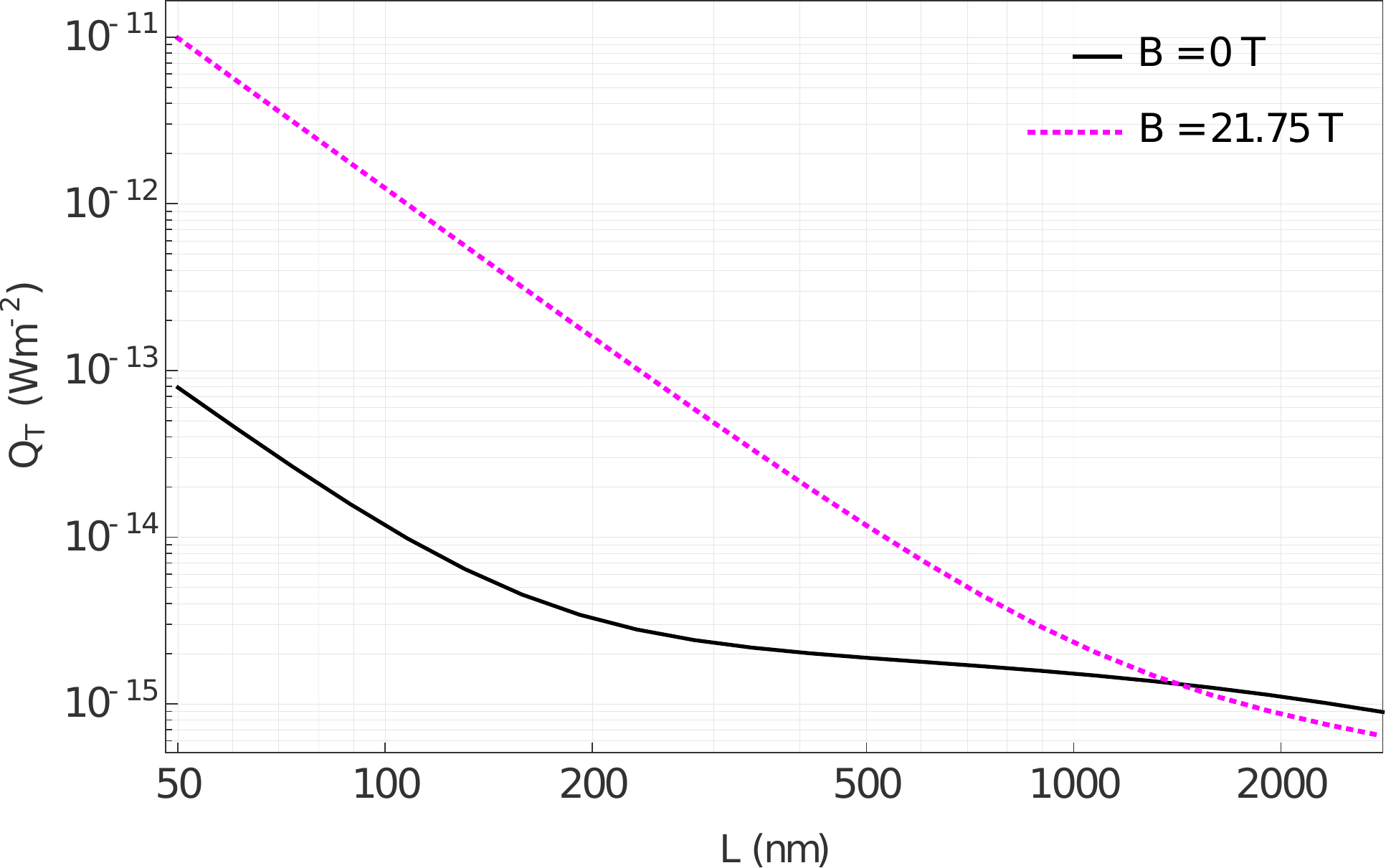}
	\caption{Total heat flux, $Q_T$, as a function of the separation distance, $L$ without magnetic field (black line) and at resonance with an applied magnetic field of $B=21.75$T (dashed-line). At short separations, there is a difference of two-orders of magnitude in the heat-flux.}
	\label{fig-06}
\end{figure}

Finally, in Fig.~\ref{fig-06}, we analyze the total heat flux, $Q_T$, as the separation distance, $L$, between the nanoparticle and the substrate increases, for the cases $B=0.0$ and $B=21.75$ T. 
We observe that the near-field heat transfer in the system increases 100 times for the short distance range $\approx 50-200$ nm when an external magnetic field of $B=21.75$ T is applied. This enhancement is progressively lost as the width of the vacuum gap exceeds the nanometer scale.

\section{Conclusions}

In this work, we presented  a theoretical study of the near-field radiative heat transfer between a magneto-optical nanoparticle and a polar surface when a static magnetic field is applied.
The magnetic field drastically modifies the polarizability of the nanoparticle: in the absence of the field, the polarizability exhibits resonances associated with phonon polaritons and surface plasmons, each of which splits into two additional satellites resonances when the magnetic field is introduced. The characteristic frequency of the new resonances arising from the plasmonic mode depends directly on the magnitude of the field. 
The resulting magneto-induced modes are no longer surface ones but become hyperbolic modes, and as the field becomes stronger, both modes available for heat transfer are progressively destroyed.
However, we found that by applying an optimal magnetic field, the total heat flux between the InSb nanoparticle and the SiC substrate increases by two orders of magnitude when the maximum polarizability of the nanoparticle is tuned to the resonance frequency of the SiC phonon-polariton.
This resonant behavior can be achieved by using another combination of magneto-optical nanoparticle and substrate with surface polaritons as long as they are located within the frequency range of magneto-optical activity.

\begin{acknowledgments}
S.~G.~C-L acknowledges support from CONACyT-Grant A1-S-10537.  
R.~E-S acknowledges partial support from DGAPA-UNAM grant IN110-819. 
\end{acknowledgments}
%
%

\begin{thebibliography}{50}
\expandafter\ifx\csname natexlab\endcsname\relax\def\natexlab#1{#1}\fi
\expandafter\ifx\csname bibnamefont\endcsname\relax
  \def\bibnamefont#1{#1}\fi
\expandafter\ifx\csname bibfnamefont\endcsname\relax
  \def\bibfnamefont#1{#1}\fi
\expandafter\ifx\csname citenamefont\endcsname\relax
  \def\citenamefont#1{#1}\fi
\expandafter\ifx\csname url\endcsname\relax
  \def\url#1{\texttt{#1}}\fi
\expandafter\ifx\csname urlprefix\endcsname\relax\def\urlprefix{URL }\fi
\providecommand{\bibinfo}[2]{#2}
\providecommand{\eprint}[2][]{\url{#2}}

\bibitem[{\citenamefont{Ben-Abdallah and Biehs}(2014)}]{PhysRevLett.112.044301}
\bibinfo{author}{\bibfnamefont{P.}~\bibnamefont{Ben-Abdallah}}
  \bibnamefont{and} \bibinfo{author}{\bibfnamefont{S.-A.} \bibnamefont{Biehs}},
  \bibinfo{journal}{Phys. Rev. Lett.} \textbf{\bibinfo{volume}{112}},
  \bibinfo{pages}{044301} (\bibinfo{year}{2014}),
  \urlprefix\url{https://link.aps.org/doi/10.1103/PhysRevLett.112.044301}.

\bibitem[{\citenamefont{DeSutter et~al.}(2019)\citenamefont{DeSutter, Tang, and
  Francoeur}}]{desutter2019near}
\bibinfo{author}{\bibfnamefont{J.}~\bibnamefont{DeSutter}},
  \bibinfo{author}{\bibfnamefont{L.}~\bibnamefont{Tang}}, \bibnamefont{and}
  \bibinfo{author}{\bibfnamefont{M.}~\bibnamefont{Francoeur}},
  \bibinfo{journal}{Nature Nano} \textbf{\bibinfo{volume}{14}},
  \bibinfo{pages}{751} (\bibinfo{year}{2019}).

\bibitem[{\citenamefont{Marconot et~al.}(2021)\citenamefont{Marconot,
  Juneau-Fecteau, and Fr{\'e}chette}}]{marconot2021toward}
\bibinfo{author}{\bibfnamefont{O.}~\bibnamefont{Marconot}},
  \bibinfo{author}{\bibfnamefont{A.}~\bibnamefont{Juneau-Fecteau}},
  \bibnamefont{and} \bibinfo{author}{\bibfnamefont{L.~G.}
  \bibnamefont{Fr{\'e}chette}}, \bibinfo{journal}{Sci. Rep.}
  \textbf{\bibinfo{volume}{11}}, \bibinfo{pages}{1} (\bibinfo{year}{2021}).

\bibitem[{\citenamefont{Kittel et~al.}(2005{\natexlab{a}})\citenamefont{Kittel,
  M\"uller-Hirsch, Parisi, Biehs, Reddig, and Holthaus}}]{Kittel05}
\bibinfo{author}{\bibfnamefont{A.}~\bibnamefont{Kittel}},
  \bibinfo{author}{\bibfnamefont{W.}~\bibnamefont{M\"uller-Hirsch}},
  \bibinfo{author}{\bibfnamefont{J.}~\bibnamefont{Parisi}},
  \bibinfo{author}{\bibfnamefont{S.-A.} \bibnamefont{Biehs}},
  \bibinfo{author}{\bibfnamefont{D.}~\bibnamefont{Reddig}}, \bibnamefont{and}
  \bibinfo{author}{\bibfnamefont{M.}~\bibnamefont{Holthaus}},
  \bibinfo{journal}{Phys. Rev. Lett.} \textbf{\bibinfo{volume}{95}},
  \bibinfo{pages}{224301} (\bibinfo{year}{2005}{\natexlab{a}}),
  \urlprefix\url{https://link.aps.org/doi/10.1103/PhysRevLett.95.224301}.

\bibitem[{\citenamefont{Francoeur}(2015)}]{Francoeur15}
\bibinfo{author}{\bibfnamefont{M.}~\bibnamefont{Francoeur}},
  \bibinfo{journal}{Nat Nano} \textbf{\bibinfo{volume}{10}},
  \bibinfo{pages}{206} (\bibinfo{year}{2015}).

\bibitem[{\citenamefont{De~Wilde et~al.}(2006)\citenamefont{De~Wilde, Formanek,
  Carminati, Gralak, Lemoine, Joulain, Mulet, Chen, and Greffet}}]{Dewilde}
\bibinfo{author}{\bibfnamefont{Y.}~\bibnamefont{De~Wilde}},
  \bibinfo{author}{\bibfnamefont{F.}~\bibnamefont{Formanek}},
  \bibinfo{author}{\bibfnamefont{R.}~\bibnamefont{Carminati}},
  \bibinfo{author}{\bibfnamefont{B.}~\bibnamefont{Gralak}},
  \bibinfo{author}{\bibfnamefont{P.-A.} \bibnamefont{Lemoine}},
  \bibinfo{author}{\bibfnamefont{K.}~\bibnamefont{Joulain}},
  \bibinfo{author}{\bibfnamefont{J.-P.} \bibnamefont{Mulet}},
  \bibinfo{author}{\bibfnamefont{Y.}~\bibnamefont{Chen}}, \bibnamefont{and}
  \bibinfo{author}{\bibfnamefont{J.-J.} \bibnamefont{Greffet}},
  \bibinfo{journal}{Nature} \textbf{\bibinfo{volume}{444}}, \bibinfo{pages}{740
  EP } (\bibinfo{year}{2006}),
  \urlprefix\url{http://dx.doi.org/10.1038/nature05265}.

\bibitem[{\citenamefont{St-Gelais et~al.}(2014)\citenamefont{St-Gelais, Guha,
  Zhu, Fan, and Lipson}}]{Gelais}
\bibinfo{author}{\bibfnamefont{R.}~\bibnamefont{St-Gelais}},
  \bibinfo{author}{\bibfnamefont{B.}~\bibnamefont{Guha}},
  \bibinfo{author}{\bibfnamefont{L.}~\bibnamefont{Zhu}},
  \bibinfo{author}{\bibfnamefont{S.}~\bibnamefont{Fan}}, \bibnamefont{and}
  \bibinfo{author}{\bibfnamefont{M.}~\bibnamefont{Lipson}},
  \bibinfo{journal}{Nano Lett.} \textbf{\bibinfo{volume}{14}},
  \bibinfo{pages}{6971} (\bibinfo{year}{2014}).

\bibitem[{\citenamefont{Kittel et~al.}(2005{\natexlab{b}})\citenamefont{Kittel,
  M\"uller-Hirsch, Parisi, Biehs, Reddig, and Holthaus}}]{kittel}
\bibinfo{author}{\bibfnamefont{A.}~\bibnamefont{Kittel}},
  \bibinfo{author}{\bibfnamefont{W.}~\bibnamefont{M\"uller-Hirsch}},
  \bibinfo{author}{\bibfnamefont{J.}~\bibnamefont{Parisi}},
  \bibinfo{author}{\bibfnamefont{S.-A.} \bibnamefont{Biehs}},
  \bibinfo{author}{\bibfnamefont{D.}~\bibnamefont{Reddig}}, \bibnamefont{and}
  \bibinfo{author}{\bibfnamefont{M.}~\bibnamefont{Holthaus}},
  \bibinfo{journal}{Phys. Rev. Lett.} \textbf{\bibinfo{volume}{95}},
  \bibinfo{pages}{224301} (\bibinfo{year}{2005}{\natexlab{b}}).

\bibitem[{\citenamefont{Vinogradov and Dorofeev}(2009)}]{Vinogradov}
\bibinfo{author}{\bibfnamefont{E.~A.} \bibnamefont{Vinogradov}}
  \bibnamefont{and} \bibinfo{author}{\bibfnamefont{I.~A.}
  \bibnamefont{Dorofeev}}, \bibinfo{journal}{Physics-Uspekhi}
  \textbf{\bibinfo{volume}{52}}, \bibinfo{pages}{425} (\bibinfo{year}{2009}).

\bibitem[{\citenamefont{Polder and Hove}(1971)}]{vanhove}
\bibinfo{author}{\bibfnamefont{D.}~\bibnamefont{Polder}} \bibnamefont{and}
  \bibinfo{author}{\bibfnamefont{M.~V.} \bibnamefont{Hove}},
  \bibinfo{journal}{Phys. Rev. B} \textbf{\bibinfo{volume}{4}},
  \bibinfo{pages}{3303} (\bibinfo{year}{1971}).

\bibitem[{\citenamefont{Hargreaves}(1969)}]{HARGREAVES}
\bibinfo{author}{\bibfnamefont{C.}~\bibnamefont{Hargreaves}},
  \bibinfo{journal}{Phys. Lett. A} \textbf{\bibinfo{volume}{30}},
  \bibinfo{pages}{491 } (\bibinfo{year}{1969}), ISSN \bibinfo{issn}{0375-9601}.

\bibitem[{\citenamefont{Santiago et~al.}(2017)\citenamefont{Santiago,
  Perez-Rodriguez, and Esquivel-Sirvent}}]{santiago2017dispersive}
\bibinfo{author}{\bibfnamefont{E.~Y.} \bibnamefont{Santiago}},
  \bibinfo{author}{\bibfnamefont{J.}~\bibnamefont{Perez-Rodriguez}},
  \bibnamefont{and}
  \bibinfo{author}{\bibfnamefont{R.}~\bibnamefont{Esquivel-Sirvent}},
  \bibinfo{journal}{J. Phys. Chem. C} \textbf{\bibinfo{volume}{121}},
  \bibinfo{pages}{12392} (\bibinfo{year}{2017}).

\bibitem[{\citenamefont{Santiago and
  Esquivel-Sirvent}(2017)}]{santiago2017near}
\bibinfo{author}{\bibfnamefont{E.~Y.} \bibnamefont{Santiago}} \bibnamefont{and}
  \bibinfo{author}{\bibfnamefont{R.}~\bibnamefont{Esquivel-Sirvent}},
  \bibinfo{journal}{Z. Naturforsch A} \textbf{\bibinfo{volume}{72}},
  \bibinfo{pages}{129} (\bibinfo{year}{2017}).

\bibitem[{\citenamefont{Biehs et~al.}(2007)\citenamefont{Biehs, Reddig, and
  Holthaus}}]{biehs2007thermal}
\bibinfo{author}{\bibfnamefont{S.-A.} \bibnamefont{Biehs}},
  \bibinfo{author}{\bibfnamefont{D.}~\bibnamefont{Reddig}}, \bibnamefont{and}
  \bibinfo{author}{\bibfnamefont{M.}~\bibnamefont{Holthaus}},
  \bibinfo{journal}{Eur. Phys. J. B} \textbf{\bibinfo{volume}{55}},
  \bibinfo{pages}{237} (\bibinfo{year}{2007}).

\bibitem[{\citenamefont{Ben-Abdallah et~al.}(2009)\citenamefont{Ben-Abdallah,
  Joulain, Drevillon, and Domingues}}]{ben2009near}
\bibinfo{author}{\bibfnamefont{P.}~\bibnamefont{Ben-Abdallah}},
  \bibinfo{author}{\bibfnamefont{K.}~\bibnamefont{Joulain}},
  \bibinfo{author}{\bibfnamefont{J.}~\bibnamefont{Drevillon}},
  \bibnamefont{and}
  \bibinfo{author}{\bibfnamefont{G.}~\bibnamefont{Domingues}},
  \bibinfo{journal}{J. App. Phys.} \textbf{\bibinfo{volume}{106}},
  \bibinfo{pages}{044306} (\bibinfo{year}{2009}).

\bibitem[{\citenamefont{Esquivel-Sirvent}(2016)}]{Thinfilmraul}
\bibinfo{author}{\bibfnamefont{R.}~\bibnamefont{Esquivel-Sirvent}},
  \bibinfo{journal}{AIP Adv.} \textbf{\bibinfo{volume}{6}},
  \bibinfo{pages}{095214} (\bibinfo{year}{2016}).

\bibitem[{\citenamefont{P{\'e}rez-Rodr{\'\i}guez
  et~al.}(2019)\citenamefont{P{\'e}rez-Rodr{\'\i}guez, Pirruccio, and
  Esquivel-Sirvent}}]{JaimeACS}
\bibinfo{author}{\bibfnamefont{J.~E.} \bibnamefont{P{\'e}rez-Rodr{\'\i}guez}},
  \bibinfo{author}{\bibfnamefont{G.}~\bibnamefont{Pirruccio}},
  \bibnamefont{and}
  \bibinfo{author}{\bibfnamefont{R.}~\bibnamefont{Esquivel-Sirvent}},
  \bibinfo{journal}{J. Phys. Chem. C} \textbf{\bibinfo{volume}{123}},
  \bibinfo{pages}{10598} (\bibinfo{year}{2019}),
  \urlprefix\url{https://doi.org/10.1021/acs.jpcc.9b01914}.

\bibitem[{\citenamefont{Lim et~al.}(2018)\citenamefont{Lim, Song, Lee, and
  Lee}}]{lim2018tailoring}
\bibinfo{author}{\bibfnamefont{M.}~\bibnamefont{Lim}},
  \bibinfo{author}{\bibfnamefont{J.}~\bibnamefont{Song}},
  \bibinfo{author}{\bibfnamefont{S.~S.} \bibnamefont{Lee}}, \bibnamefont{and}
  \bibinfo{author}{\bibfnamefont{B.~J.} \bibnamefont{Lee}},
  \bibinfo{journal}{Nat. Comm.} \textbf{\bibinfo{volume}{9}},
  \bibinfo{pages}{1} (\bibinfo{year}{2018}).

\bibitem[{\citenamefont{P\'erez-Rodr\'{\i}guez
  et~al.}(2019)\citenamefont{P\'erez-Rodr\'{\i}guez, Pirruccio, and
  Esquivel-Sirvent}}]{PhysRevMaterials.3.015201}
\bibinfo{author}{\bibfnamefont{J.~E.} \bibnamefont{P\'erez-Rodr\'{\i}guez}},
  \bibinfo{author}{\bibfnamefont{G.}~\bibnamefont{Pirruccio}},
  \bibnamefont{and}
  \bibinfo{author}{\bibfnamefont{R.}~\bibnamefont{Esquivel-Sirvent}},
  \bibinfo{journal}{Phys. Rev. Materials} \textbf{\bibinfo{volume}{3}},
  \bibinfo{pages}{015201} (\bibinfo{year}{2019}),
  \urlprefix\url{https://link.aps.org/doi/10.1103/PhysRevMaterials.3.015201}.

\bibitem[{\citenamefont{P\'erez-Rodr\'{\i}guez
  et~al.}(2017)\citenamefont{P\'erez-Rodr\'{\i}guez, Pirruccio, and
  Esquivel-Sirvent}}]{Esquivel17}
\bibinfo{author}{\bibfnamefont{J.~E.} \bibnamefont{P\'erez-Rodr\'{\i}guez}},
  \bibinfo{author}{\bibfnamefont{G.}~\bibnamefont{Pirruccio}},
  \bibnamefont{and}
  \bibinfo{author}{\bibfnamefont{R.}~\bibnamefont{Esquivel-Sirvent}},
  \bibinfo{journal}{Phys. Rev. Mat.} \textbf{\bibinfo{volume}{1}},
  \bibinfo{pages}{062201} (\bibinfo{year}{2017}),
  \urlprefix\url{https://link.aps.org/doi/10.1103/PhysRevMaterials.1.062201}.

\bibitem[{\citenamefont{Ghanekar et~al.}(2018)\citenamefont{Ghanekar, Tian,
  Ricci, Zhang, Gregory, and Zheng}}]{ghanekar2018near}
\bibinfo{author}{\bibfnamefont{A.}~\bibnamefont{Ghanekar}},
  \bibinfo{author}{\bibfnamefont{Y.}~\bibnamefont{Tian}},
  \bibinfo{author}{\bibfnamefont{M.}~\bibnamefont{Ricci}},
  \bibinfo{author}{\bibfnamefont{S.}~\bibnamefont{Zhang}},
  \bibinfo{author}{\bibfnamefont{O.}~\bibnamefont{Gregory}}, \bibnamefont{and}
  \bibinfo{author}{\bibfnamefont{Y.}~\bibnamefont{Zheng}},
  \bibinfo{journal}{Opt. Exp.} \textbf{\bibinfo{volume}{26}},
  \bibinfo{pages}{A209} (\bibinfo{year}{2018}).

\bibitem[{\citenamefont{Castillo-L{\'o}pez
  et~al.}(2020)\citenamefont{Castillo-L{\'o}pez, Pirruccio, Villarreal, and
  Esquivel-Sirvent}}]{castillo2020near}
\bibinfo{author}{\bibfnamefont{S.}~\bibnamefont{Castillo-L{\'o}pez}},
  \bibinfo{author}{\bibfnamefont{G.}~\bibnamefont{Pirruccio}},
  \bibinfo{author}{\bibfnamefont{C.}~\bibnamefont{Villarreal}},
  \bibnamefont{and}
  \bibinfo{author}{\bibfnamefont{R.}~\bibnamefont{Esquivel-Sirvent}},
  \bibinfo{journal}{Sci. Rep.} \textbf{\bibinfo{volume}{10}},
  \bibinfo{pages}{1} (\bibinfo{year}{2020}).

\bibitem[{\citenamefont{Castillo-L{\'o}pez
  et~al.}(2022)\citenamefont{Castillo-L{\'o}pez, Villarreal, Esquivel-Sirvent,
  and Pirruccio}}]{castillo2022enhancing}
\bibinfo{author}{\bibfnamefont{S.}~\bibnamefont{Castillo-L{\'o}pez}},
  \bibinfo{author}{\bibfnamefont{C.}~\bibnamefont{Villarreal}},
  \bibinfo{author}{\bibfnamefont{R.}~\bibnamefont{Esquivel-Sirvent}},
  \bibnamefont{and}
  \bibinfo{author}{\bibfnamefont{G.}~\bibnamefont{Pirruccio}},
  \bibinfo{journal}{Int. J. Heat Mass Transf.} \textbf{\bibinfo{volume}{182}},
  \bibinfo{pages}{121922} (\bibinfo{year}{2022}).

\bibitem[{\citenamefont{Becerril and Noguez}(2019)}]{PhysRevB.99.045418}
\bibinfo{author}{\bibfnamefont{D.}~\bibnamefont{Becerril}} \bibnamefont{and}
  \bibinfo{author}{\bibfnamefont{C.}~\bibnamefont{Noguez}},
  \bibinfo{journal}{Phys. Rev. B} \textbf{\bibinfo{volume}{99}},
  \bibinfo{pages}{045418} (\bibinfo{year}{2019}),
  \urlprefix\url{https://link.aps.org/doi/10.1103/PhysRevB.99.045418}.

\bibitem[{\citenamefont{Wang and Wu}(2016)}]{Wang16}
\bibinfo{author}{\bibfnamefont{Y.}~\bibnamefont{Wang}} \bibnamefont{and}
  \bibinfo{author}{\bibfnamefont{J.}~\bibnamefont{Wu}}, \bibinfo{journal}{AIP
  Adv.} \textbf{\bibinfo{volume}{6}}, \bibinfo{pages}{025104}
  (\bibinfo{year}{2016}), \eprint{https://doi.org/10.1063/1.4941751},
  \urlprefix\url{https://doi.org/10.1063/1.4941751}.

\bibitem[{\citenamefont{Chapuis
  et~al.}(2008{\natexlab{a}})\citenamefont{Chapuis, Laroche, Volz, and
  Greffet}}]{chapuis2008radiative}
\bibinfo{author}{\bibfnamefont{P.-O.} \bibnamefont{Chapuis}},
  \bibinfo{author}{\bibfnamefont{M.}~\bibnamefont{Laroche}},
  \bibinfo{author}{\bibfnamefont{S.}~\bibnamefont{Volz}}, \bibnamefont{and}
  \bibinfo{author}{\bibfnamefont{J.-J.} \bibnamefont{Greffet}},
  \bibinfo{journal}{App. Phys. Lett.} \textbf{\bibinfo{volume}{92}},
  \bibinfo{pages}{201906} (\bibinfo{year}{2008}{\natexlab{a}}).

\bibitem[{\citenamefont{Bai et~al.}(2015)\citenamefont{Bai, Jiang, and
  Liu}}]{bai2015enhanced}
\bibinfo{author}{\bibfnamefont{Y.}~\bibnamefont{Bai}},
  \bibinfo{author}{\bibfnamefont{Y.}~\bibnamefont{Jiang}}, \bibnamefont{and}
  \bibinfo{author}{\bibfnamefont{L.}~\bibnamefont{Liu}}, \bibinfo{journal}{J.
  Quant. Spect. Rad. Transf.} \textbf{\bibinfo{volume}{158}},
  \bibinfo{pages}{61} (\bibinfo{year}{2015}).

\bibitem[{\citenamefont{Huth et~al.}(2010)\citenamefont{Huth, R{\"u}ting,
  Biehs, and Holthaus}}]{huth2010shape}
\bibinfo{author}{\bibfnamefont{O.}~\bibnamefont{Huth}},
  \bibinfo{author}{\bibfnamefont{F.}~\bibnamefont{R{\"u}ting}},
  \bibinfo{author}{\bibfnamefont{S.-A.} \bibnamefont{Biehs}}, \bibnamefont{and}
  \bibinfo{author}{\bibfnamefont{M.}~\bibnamefont{Holthaus}},
  \bibinfo{journal}{Eur. Phys. J. -App. Phys.} \textbf{\bibinfo{volume}{50}}
  (\bibinfo{year}{2010}).

\bibitem[{\citenamefont{Barnes}(2016)}]{shape1}
\bibinfo{author}{\bibfnamefont{W.~L.} \bibnamefont{Barnes}},
  \bibinfo{journal}{Am. J. Phys.} \textbf{\bibinfo{volume}{84}},
  \bibinfo{pages}{593} (\bibinfo{year}{2016}).

\bibitem[{\citenamefont{Noguez}(2007)}]{Cecilia}
\bibinfo{author}{\bibfnamefont{C.}~\bibnamefont{Noguez}}, \bibinfo{journal}{J.
  Phys. Chem. C} \textbf{\bibinfo{volume}{111}}, \bibinfo{pages}{3806}
  (\bibinfo{year}{2007}), \urlprefix\url{https://doi.org/10.1021/jp066539m}.

\bibitem[{\citenamefont{Kushwaha}(2001)}]{KUSHWAHA20011}
\bibinfo{author}{\bibfnamefont{M.~S.} \bibnamefont{Kushwaha}},
  \bibinfo{journal}{Surf. Sci. Rep.} \textbf{\bibinfo{volume}{41}},
  \bibinfo{pages}{1 } (\bibinfo{year}{2001}), ISSN \bibinfo{issn}{0167-5729},
  \urlprefix\url{http://www.sciencedirect.com/science/article/pii/S0167572900000078}.

\bibitem[{\citenamefont{Keyes et~al.}(1956)\citenamefont{Keyes, Zwerdling,
  Foner, Kolm, and Lax}}]{keyes1956infrared}
\bibinfo{author}{\bibfnamefont{R.~J.} \bibnamefont{Keyes}},
  \bibinfo{author}{\bibfnamefont{S.}~\bibnamefont{Zwerdling}},
  \bibinfo{author}{\bibfnamefont{S.}~\bibnamefont{Foner}},
  \bibinfo{author}{\bibfnamefont{H.}~\bibnamefont{Kolm}}, \bibnamefont{and}
  \bibinfo{author}{\bibfnamefont{B.}~\bibnamefont{Lax}},
  \bibinfo{journal}{Phys. Rev.} \textbf{\bibinfo{volume}{104}},
  \bibinfo{pages}{1804} (\bibinfo{year}{1956}).

\bibitem[{\citenamefont{Palik et~al.}(1961)\citenamefont{Palik, Picus, Teitler,
  and Wallis}}]{palik1961infrared}
\bibinfo{author}{\bibfnamefont{E.}~\bibnamefont{Palik}},
  \bibinfo{author}{\bibfnamefont{G.}~\bibnamefont{Picus}},
  \bibinfo{author}{\bibfnamefont{S.}~\bibnamefont{Teitler}}, \bibnamefont{and}
  \bibinfo{author}{\bibfnamefont{R.}~\bibnamefont{Wallis}},
  \bibinfo{journal}{Phys. Rev.} \textbf{\bibinfo{volume}{122}},
  \bibinfo{pages}{475} (\bibinfo{year}{1961}).

\bibitem[{\citenamefont{Palik and Furdyna}(1970)}]{Palik_1970}
\bibinfo{author}{\bibfnamefont{E.~D.} \bibnamefont{Palik}} \bibnamefont{and}
  \bibinfo{author}{\bibfnamefont{J.~K.} \bibnamefont{Furdyna}},
  \bibinfo{journal}{Rep. Prog. Phys.} \textbf{\bibinfo{volume}{33}},
  \bibinfo{pages}{1193} (\bibinfo{year}{1970}),
  \urlprefix\url{https://doi.org/10.1088%2F0034-4885%2F33%2F3%2F307}.

\bibitem[{\citenamefont{Chochol et~al.}(2016)\citenamefont{Chochol, Postava,
  {\v{C}}ada, Vanwolleghem, Halaga{\v{c}}ka, Lampin, and
  Pi{\v{s}}tora}}]{chochol2016magneto}
\bibinfo{author}{\bibfnamefont{J.}~\bibnamefont{Chochol}},
  \bibinfo{author}{\bibfnamefont{K.}~\bibnamefont{Postava}},
  \bibinfo{author}{\bibfnamefont{M.}~\bibnamefont{{\v{C}}ada}},
  \bibinfo{author}{\bibfnamefont{M.}~\bibnamefont{Vanwolleghem}},
  \bibinfo{author}{\bibfnamefont{L.}~\bibnamefont{Halaga{\v{c}}ka}},
  \bibinfo{author}{\bibfnamefont{J.-F.} \bibnamefont{Lampin}},
  \bibnamefont{and}
  \bibinfo{author}{\bibfnamefont{J.}~\bibnamefont{Pi{\v{s}}tora}},
  \bibinfo{journal}{AIP Adv.} \textbf{\bibinfo{volume}{6}},
  \bibinfo{pages}{115021} (\bibinfo{year}{2016}).

\bibitem[{\citenamefont{Dragoman and Dragoman}(2008)}]{dragoman2008plasmonics}
\bibinfo{author}{\bibfnamefont{M.}~\bibnamefont{Dragoman}} \bibnamefont{and}
  \bibinfo{author}{\bibfnamefont{D.}~\bibnamefont{Dragoman}},
  \bibinfo{journal}{Prog. Quant. Electron.} \textbf{\bibinfo{volume}{32}},
  \bibinfo{pages}{1} (\bibinfo{year}{2008}).

\bibitem[{\citenamefont{Pedersen}(2020)}]{PhysRevB.102.075410}
\bibinfo{author}{\bibfnamefont{T.~G.} \bibnamefont{Pedersen}},
  \bibinfo{journal}{Phys. Rev. B} \textbf{\bibinfo{volume}{102}},
  \bibinfo{pages}{075410} (\bibinfo{year}{2020}),
  \urlprefix\url{https://link.aps.org/doi/10.1103/PhysRevB.102.075410}.

\bibitem[{\citenamefont{Zhang et~al.}(2015)\citenamefont{Zhang, Ren, and
  Yao}}]{zhang2015surface}
\bibinfo{author}{\bibfnamefont{Y.~M.} \bibnamefont{Zhang}},
  \bibinfo{author}{\bibfnamefont{G.~J.} \bibnamefont{Ren}}, \bibnamefont{and}
  \bibinfo{author}{\bibfnamefont{J.~Q.} \bibnamefont{Yao}},
  \bibinfo{journal}{Opt. Comm.} \textbf{\bibinfo{volume}{341}},
  \bibinfo{pages}{173} (\bibinfo{year}{2015}).

\bibitem[{\citenamefont{Liu et~al.}(2012)\citenamefont{Liu, Chang, Schaller,
  and Talapin}}]{InSbparticle}
\bibinfo{author}{\bibfnamefont{W.}~\bibnamefont{Liu}},
  \bibinfo{author}{\bibfnamefont{A.~Y.} \bibnamefont{Chang}},
  \bibinfo{author}{\bibfnamefont{R.~D.} \bibnamefont{Schaller}},
  \bibnamefont{and} \bibinfo{author}{\bibfnamefont{D.~V.}
  \bibnamefont{Talapin}}, \bibinfo{journal}{J. Am. Chem. Soc.}
  \textbf{\bibinfo{volume}{134}}, \bibinfo{pages}{20258}
  (\bibinfo{year}{2012}), \urlprefix\url{https://doi.org/10.1021/ja309821j}.

\bibitem[{\citenamefont{Abraham~Ekeroth
  et~al.}(2018)\citenamefont{Abraham~Ekeroth, Ben-Abdallah, Cuevas, and
  Garc{\'\i}a-Mart{\'\i}n}}]{abraham2018anisotropic}
\bibinfo{author}{\bibfnamefont{R.~M.} \bibnamefont{Abraham~Ekeroth}},
  \bibinfo{author}{\bibfnamefont{P.}~\bibnamefont{Ben-Abdallah}},
  \bibinfo{author}{\bibfnamefont{J.~C.} \bibnamefont{Cuevas}},
  \bibnamefont{and}
  \bibinfo{author}{\bibfnamefont{A.}~\bibnamefont{Garc{\'\i}a-Mart{\'\i}n}},
  \bibinfo{journal}{ACS Phot.} \textbf{\bibinfo{volume}{5}},
  \bibinfo{pages}{705} (\bibinfo{year}{2018}).

\bibitem[{\citenamefont{Latella and
  Ben-Abdallah}(2017)}]{PhysRevLett.118.173902}
\bibinfo{author}{\bibfnamefont{I.}~\bibnamefont{Latella}} \bibnamefont{and}
  \bibinfo{author}{\bibfnamefont{P.}~\bibnamefont{Ben-Abdallah}},
  \bibinfo{journal}{Phys. Rev. Lett.} \textbf{\bibinfo{volume}{118}},
  \bibinfo{pages}{173902} (\bibinfo{year}{2017}),
  \urlprefix\url{https://link.aps.org/doi/10.1103/PhysRevLett.118.173902}.

\bibitem[{\citenamefont{M\'{a}rquez and Esquivel-Sirvent}(2020)}]{Marquez:20}
\bibinfo{author}{\bibfnamefont{A.}~\bibnamefont{M\'{a}rquez}} \bibnamefont{and}
  \bibinfo{author}{\bibfnamefont{R.}~\bibnamefont{Esquivel-Sirvent}},
  \bibinfo{journal}{Opt. Express} \textbf{\bibinfo{volume}{28}},
  \bibinfo{pages}{39005} (\bibinfo{year}{2020}),
  \urlprefix\url{http://www.osapublishing.org/oe/abstract.cfm?URI=oe-28-26-39005}.

\bibitem[{\citenamefont{Shuvaev et~al.}(2021)\citenamefont{Shuvaev, Muravev,
  Gusikhin, Gospodari\ifmmode~\check{c}\else \v{c}\fi{}, Pimenov, and
  Kukushkin}}]{PhysRevLett.126.136801}
\bibinfo{author}{\bibfnamefont{A.}~\bibnamefont{Shuvaev}},
  \bibinfo{author}{\bibfnamefont{V.~M.} \bibnamefont{Muravev}},
  \bibinfo{author}{\bibfnamefont{P.~A.} \bibnamefont{Gusikhin}},
  \bibinfo{author}{\bibfnamefont{J.}~\bibnamefont{Gospodari\ifmmode~\check{c}\else
  \v{c}\fi{}}}, \bibinfo{author}{\bibfnamefont{A.}~\bibnamefont{Pimenov}},
  \bibnamefont{and} \bibinfo{author}{\bibfnamefont{I.~V.}
  \bibnamefont{Kukushkin}}, \bibinfo{journal}{Phys. Rev. Lett.}
  \textbf{\bibinfo{volume}{126}}, \bibinfo{pages}{136801}
  (\bibinfo{year}{2021}),
  \urlprefix\url{https://link.aps.org/doi/10.1103/PhysRevLett.126.136801}.

\bibitem[{\citenamefont{Pandya and Kordesch}(2015)}]{kordesch}
\bibinfo{author}{\bibfnamefont{S.~G.} \bibnamefont{Pandya}} \bibnamefont{and}
  \bibinfo{author}{\bibfnamefont{M.~E.} \bibnamefont{Kordesch}},
  \bibinfo{journal}{Nanoscale Res. Lett.} \textbf{\bibinfo{volume}{10}},
  \bibinfo{pages}{258} (\bibinfo{year}{2015}),
  \urlprefix\url{https://doi.org/10.1186/s11671-015-0966-4}.

\bibitem[{\citenamefont{Hartstein et~al.}(1975)\citenamefont{Hartstein,
  Burstein, Palik, Gammon, and Henvis}}]{PhysRevB.12.3186}
\bibinfo{author}{\bibfnamefont{A.}~\bibnamefont{Hartstein}},
  \bibinfo{author}{\bibfnamefont{E.}~\bibnamefont{Burstein}},
  \bibinfo{author}{\bibfnamefont{E.~D.} \bibnamefont{Palik}},
  \bibinfo{author}{\bibfnamefont{R.~W.} \bibnamefont{Gammon}},
  \bibnamefont{and} \bibinfo{author}{\bibfnamefont{B.~W.}
  \bibnamefont{Henvis}}, \bibinfo{journal}{Phys. Rev. B}
  \textbf{\bibinfo{volume}{12}}, \bibinfo{pages}{3186} (\bibinfo{year}{1975}),
  \urlprefix\url{https://link.aps.org/doi/10.1103/PhysRevB.12.3186}.

\bibitem[{\citenamefont{Weick and Weinmann}(2011)}]{Weick}
\bibinfo{author}{\bibfnamefont{G.}~\bibnamefont{Weick}} \bibnamefont{and}
  \bibinfo{author}{\bibfnamefont{D.}~\bibnamefont{Weinmann}},
  \bibinfo{journal}{Phys. Rev. B} \textbf{\bibinfo{volume}{83}},
  \bibinfo{pages}{125405} (\bibinfo{year}{2011}),
  \urlprefix\url{https://link.aps.org/doi/10.1103/PhysRevB.83.125405}.

\bibitem[{\citenamefont{Chapuis
  et~al.}(2008{\natexlab{b}})\citenamefont{Chapuis, Laroche, Volz, and
  Greffet}}]{PhysRevB.77.125402}
\bibinfo{author}{\bibfnamefont{P.-O.} \bibnamefont{Chapuis}},
  \bibinfo{author}{\bibfnamefont{M.}~\bibnamefont{Laroche}},
  \bibinfo{author}{\bibfnamefont{S.}~\bibnamefont{Volz}}, \bibnamefont{and}
  \bibinfo{author}{\bibfnamefont{J.-J.} \bibnamefont{Greffet}},
  \bibinfo{journal}{Phys. Rev. B} \textbf{\bibinfo{volume}{77}},
  \bibinfo{pages}{125402} (\bibinfo{year}{2008}{\natexlab{b}}),
  \urlprefix\url{https://link.aps.org/doi/10.1103/PhysRevB.77.125402}.

\bibitem[{\citenamefont{Bohren and Huffman}(2008)}]{bohren2008absorption}
\bibinfo{author}{\bibfnamefont{C.~F.} \bibnamefont{Bohren}} \bibnamefont{and}
  \bibinfo{author}{\bibfnamefont{D.~R.} \bibnamefont{Huffman}},
  \emph{\bibinfo{title}{Absorption and scattering of light by small particles}}
  (\bibinfo{publisher}{John Wiley \& Sons}, \bibinfo{year}{2008}).

\bibitem[{\citenamefont{Hong et~al.}(2020)\citenamefont{Hong, Siahmakoun, and
  Alisafaee}}]{hong2020biaxial}
\bibinfo{author}{\bibfnamefont{C.}~\bibnamefont{Hong}},
  \bibinfo{author}{\bibfnamefont{A.}~\bibnamefont{Siahmakoun}},
  \bibnamefont{and}
  \bibinfo{author}{\bibfnamefont{H.}~\bibnamefont{Alisafaee}}, in
  \emph{\bibinfo{booktitle}{Photonic and Phononic Properties of Engineered
  Nanostructures X}} (\bibinfo{organization}{International Society for Optics
  and Photonics}, \bibinfo{year}{2020}), vol. \bibinfo{volume}{11289}, p.
  \bibinfo{pages}{1128929}.

\bibitem[{\citenamefont{Song et~al.}(2018)\citenamefont{Song, Liu, Xiang, and
  Aydin}}]{song2018biaxial}
\bibinfo{author}{\bibfnamefont{X.}~\bibnamefont{Song}},
  \bibinfo{author}{\bibfnamefont{Z.}~\bibnamefont{Liu}},
  \bibinfo{author}{\bibfnamefont{Y.}~\bibnamefont{Xiang}}, \bibnamefont{and}
  \bibinfo{author}{\bibfnamefont{K.}~\bibnamefont{Aydin}},
  \bibinfo{journal}{Opt. Express} \textbf{\bibinfo{volume}{26}},
  \bibinfo{pages}{5469} (\bibinfo{year}{2018}).

\end{thebibliography}
%

%

\end{document}